# Rare-earth effect on the physical properties of $Na_{0.5}Bi_{0.5}TiO_3$ system: A Density Functional Theory investigation

Manal Benyoussef[1], Halima Zaari[2], Jamal Belhadi[1], Youssef El Amraoui[2,3], Hamid Ez-Zahraouy[2], Abdelilah Lahmar[1], Mimoun El Marssi[1]

[1] Laboratory of Physics of Condensed Matter (LPMC), University of Picardie Jules Verne, Scientific Pole, 80039 Amiens Cedex 1, France
[2] Laboratory of Condensed Matter and Indisciplinary Science (LaMCScI), Faculty of Science, Mohammed V University, BP 1014, Rabat, Morocco
[3] National School of Arts and Crafts, Moulay Ismail University, 50500, Meknes, Morocco

**Abstract**

$Na_{0.5}(Bi_{3/4}RE_{1/4})_{0.5}TiO_3$ (RENBT, RE = Nd, Gd, Dy, and Ho) compounds were investigated in the framework of first-principles calculations using the full potential linearized augmented plane wave (FP-LAPW) method based on the spin-polarized density functional theory implemented in the WIEN2k code. Combined charge density distribution and Ti K-edge x-ray absorption spectra revealed that the RENBT compositions with high polarization values were accompanied by a higher $TiO_6$ distortion, DyNBT, and NdNBT compounds. The effect of the rare-earth elements on the polarization were confirmed experimentally with the collection of the hysteresis loops. The investigation of the electronic properties of the compounds highlighted the emergence of a magnetization owing to the 4f orbital effect of the rare-earth elements. Besides, the investigation of the chemical ordering showed a short-range chemical ordering for the pure composition and an increased A-site disorder for dysprosium doped NBT system. The increased disorder may speak for increased relaxor properties in the RE doped compositions.

Corresponding author: manal.benyoussef@etud.u-picardie.fr, Tel.: +33 322827837

Foundation items: Project supported by the Haute France Region/ FEDER (project MASEN) and H2020-RISE-ENGIMA-778072 project.

## 1. Introduction

Since the European Union has restricted the use of lead in the electronic equipment (2011/65/Eu (RoHS) [1] owing to environmental and health issues, research activity got oriented into lead-free materials with enhanced properties. Among the most studied lead-free ferroelectric perovskite having promising properties, one can found the A-site mixed relaxor ferroelectric $Na_{0.5}Bi_{0.5}TiO_3$ (NBT) system. Thanks to its rich structural and dielectric characteristics, the latter material was seriously considered to replace lead-based materials in power electronic applications [2–5]. It should be noted that NBT system possess a tunable defect chemistry and can be turned into an insulator for dielectric capacitors, or to an oxide-ion conductor for electrochemical devices and sensors. The mobile ionic defect present in NBT system (i.e., oxygen vacancies), can be either enhanced or decreased using specific doping elements. For instance, if acceptor doping is used, the ionic conductivity of NBT can be significantly enhanced making it an excellent oxide-ion conductor [6–10]. On the other hand, Yang *et al.* demonstrated that isovalent doping (such as $RE^{3+}$ to replace $Bi^{3+}$) resulted in an important decrease of the conductivity within NBT system which was related to the lower polarizability of the rare-earth elements [6]. In our previous experimental investigation, we demonstrated that Dy doping considerably reduces the conductivity of NBT system owing to the decreased oxygen vacancies [13]. Therefore, in the present work, we will highlight the role of the substitution of $Bi^{3+}$ by $RE^{3+}$ ions on the physical properties of the ferroelectric $Na_{0.5}Bi_{0.5}TiO_3$ system.

Besides, the correlation of the ferroelectric behaviors of perovskite materials to their structural and electronic properties is noteworthy and can be investigated using the Ti K-edge/O K-edge features using the x-ray absorption spectra [14–16]. Within this scope, correlations will be given between the occurring octahedral $TiO_6$ distortions and the ferroelectric character of the studied RENBT systems using the investigation of the x-ray absorption near edge structure (XANES) at the Ti K-edge.

Rare earth doping was reported to be an effective alternative way to improve the physical properties of ferroelectric systems [17–27]. Besides, based on lead-based ferroelectric systems, doping with RE ions was found to induce a change in the ordering degree of A/B-site cations and induce random fields/bonds that give rise to local structural heterogeneities [28–30]. These last can generate short-range polar structures exhibiting obvious relaxor characteristics with enhanced physical properties [31–35]. Since the A-site ordering degree is noteworthy in the emergence of the relaxor behaviors in A/B site mixed perovskites [36], it's thus necessary to consider the chemical ordering in the rare-earth-doped NBT systems. Several studies were interested in the investigation of the A-site chemical ordering of NBT system by both theoretical and experimental methods, however, this topic remains much debated [37–40]. The focus is generally put on the study of pure NBT system, in its high-temperature cubic structure. However, theoretical investigations on the effect of rare-earth on the physical properties of NBT system are scarce [41]. Therefore, one of the aims of this study is to investigate the A-site chemical ordering of the RENBT system, in their ferroelectric rhombohedral (R3c) structure, to evaluate the effect of $RE^{3+}$ ions on the relaxor properties of the system. The present work will give interesting results about the structural, electronic, ferroelectric, and magnetic properties of rare-earth-doped NBT systems by the use of the density functional theory. In addition, the effect of rare-earth on the ferroelectric and relaxor properties of RENBT ceramics will be confirmed experimentally.

## 2. Computational and experimental methods

All calculations were performed using the full-potential linearized augmented plane wave (FP-LAPW) method based on the spin-polarized density functional theory, as implemented in the WIEN2k code [42]. The muffin tin radii ($R_{MT}$) were taken to be 2.24 a.u. (atomic units) for Na, and 2.36, 1.76, 1.59 a.u. for Bi, Ti, and O, respectively. The $R_{MT}$ of the rare earth elements used in the framework of this study was taken as, 2.46 a.u. for neodymium, 2.40 a.u. for gadolinium, 2.37 a.u. for dysprosium, and 2.36 a.u. for holmium. The rhombohedral structure of NBT with R3c ($C_{3v}^6$) space group was taken in this study due to its ferroelectric properties. To investigate the effect of rare-earth doping on the physical properties of NBT matrix, a supercell of 2×1×2 (40 atoms) was considered. Notice that, the supercell of the parent compound contains eight A-cations (4 Na and 4 Bi) on the rhombohedral structure. To consider rare-earth doping on the structure, ¼ bismuth atom was substituted by one rare earth atom (RE = $Nd^{3+}$, $Gd^{3+}$, $Dy^{3+}$, $Ho^{3+}$). A relaxation of the atomic positions of all studied compositions was performed through the QUANTUM ESPRESSO code [43]. During such relaxation, the atoms were free to move to obtain relaxed internal forces. The force convergence was taken as $10^{-5}$ Ry/a.u. for the total energy convergence. The Perdew-Burke-Ernzerhof (PBE) [44] pseudopotentials were used for the calculations. The resulted relaxed cell parameters and atomic positions were used as input in the Wien2k code. The Berry phase PI approach [45] implemented in the Wien2k code permits us to investigate the spontaneous polarization of our studied compositions. The exchange-correlation energy is approximated within the generalized gradient approximation (GGA). We use 200 k-points for Brillouin zone integration [46]. To confirm the evolution of the polarization as a function of rare-earth doping, $Na_{0.5}(Bi_{0.98}RE_{0.02})_{0.5}TiO_3$ ceramics in the rhombohedral (R3c) phase were elaborated using the conventional solid-state method. The details of the synthesis can be found elsewhere [13,18]. The rhombohedral phase of the RENBT materials was confirmed using the Rietveld refinement of the x-ray diffraction patterns [18]. The ferroelectric hysteresis loops were collected at RT and 1Hz using a ferroelectric test system (TF Analyzer 2000, aix-ACCT).

## 3. Results and discussion
### 3.1. Crystal structure

The crystal structure of the 40 atoms supercell of NBT and rare-earth doped NBT system (RE = Nd, Gd, Dy, and Ho) are presented in Fig. S1(a, and b) (see supplementary informations). The relaxed lattice parameters of the pure and doped NBT structures are gathered in Table 1. One can see from the table, a decrease in the lattice constant, $a_0$, and unit cell volume of the doped compositions (~5.58 Å) compared to the parent NBT system (~5.60 Å). The main reason for this decrease, is the substitution of bismuth ($r_{Bi3+}$ = 1.17 Å) by a lower ionic radius, such as, $r_{Nd3+}$ = 1.109 Å, $r_{Gd3+}$ = 1.053 Å, $r_{Dy3+}$ = 1.027 Å, and $r_{Ho3+}$ = 1.015 Å.

**Table 1.** Computed lattice constant $a_0$, angle α, and unit cell volume of the compounds (R3c).

|  | **NBT** | **NdNBT** | **GdNBT** | **DyNBT** | **HoNBT** |
|---|---|---|---|---|---|
| **$a_0$ (Å)** | 5.609 | 5.582 | 5.580 | 5.584 | 5.581 |

| α (°) | 59.19 | 59.38 | 59.27 | 59.15 | 59.17 |
| V (Å³) | 122.5 | 121.25 | 120.81 | 120.73 | 120.59 |

### 3.2. Charge density distribution and ferroelectric properties

Fig. 1(a-e) present the charge density distribution of RENBT systems. The mapping shows well the strong Bi-O and Ti-O hybridization. Notice that, Bi, Ti, and RE atoms are observed to share covalent bonding with oxygen atoms, whereas Na-O has an ionic bonding. The substitution of $Bi^{3+}$ by $Nd^{3+}$ ion which present a partially filled 4f orbital ($4f^3$), resulted in enhanced Nd-O bond length and reduced the charge density for this composition.

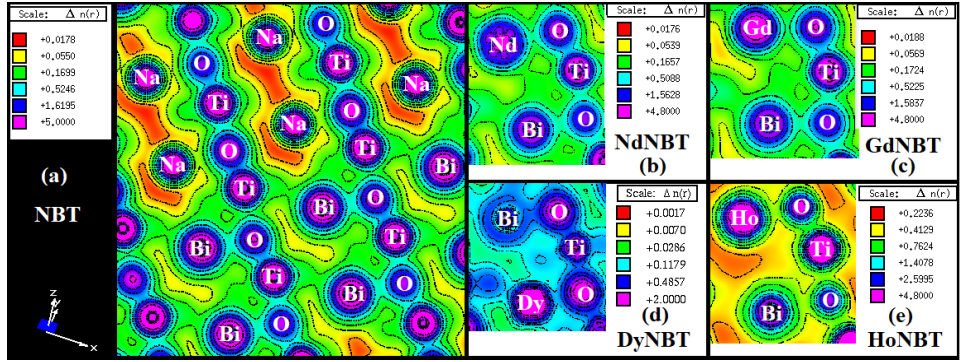

**Fig. 1:** The Charge density distribution of (a) NBT, (b) NdNBT, (c) GdNBT, (d) DyNBT, and (e) HoNBT compositions.

To have a better view of the occurring changes, we illustrated in Fig. 2(a-d) the bond length of the different atoms with the oxygen atom. Fig. 2(a) presents the RE-O bond length of the different RENBT compositions. The larger Nd-O bond length compared to the Bi-O one, resulted in a disturbance of the surrounding bismuth atoms and gave rise to a reinforced Bi-O hybridization for the NdNBT composition (Fig. 2(b)). As the ionic radius is decreasing (from 1.109 Å for Nd to 1.015 Å for Ho), the RE-O bond length is gradually decreasing, and the Bi-O hybridization is observed to present the reverse trend with a gradual increase, i.e., a weaker Bi-O hybridization. The trend can also be confirmed in Fig. 1, where a charge density distribution of less intensity is observed for Bi-O hybridization in the HoNBT system.

Notice that, the Ti-O bond length (Fig. 2(c)) of NBT, GdNBT, and HoNBT are of similar magnitude with a small decrease in the case of the doped compositions. Interestingly, a higher titanium displacement is observed for neodymium and dysprosium doped NBT systems compared to the NBT composition. Remind that Bi-O and Ti-O interactions are pertinent in the emergence of ferroelectricity in the system. Therefore, to correlate the strength of the hybridizations obtained from the charge density distribution and their corresponding bond lengths to the ferroelectric properties of the studied compounds, we computed the spontaneous polarization ($P_s$) of the RENBT materials using the Berry phase PI approach (Fig. 2(d)). As observed, a high polarization value is obtained for the NBT system (~38 µC/cm²) and matches well with the experimental and theoretical reported values [13,47]. Note that the ferroelectricity in NBT system is known to be related to the lone pair effect ($6s^2$) and the higher polarizability of bismuth atom [6]. Ideally and without chemistry defects, replacing the bismuth atom by a rare earth element (4f), presenting a lower polarizability value, will decrease the number of bismuth lone pairs, and consequently, the polarization of the doped compounds will be lower than that of the NBT system (Fig. 2(d)).

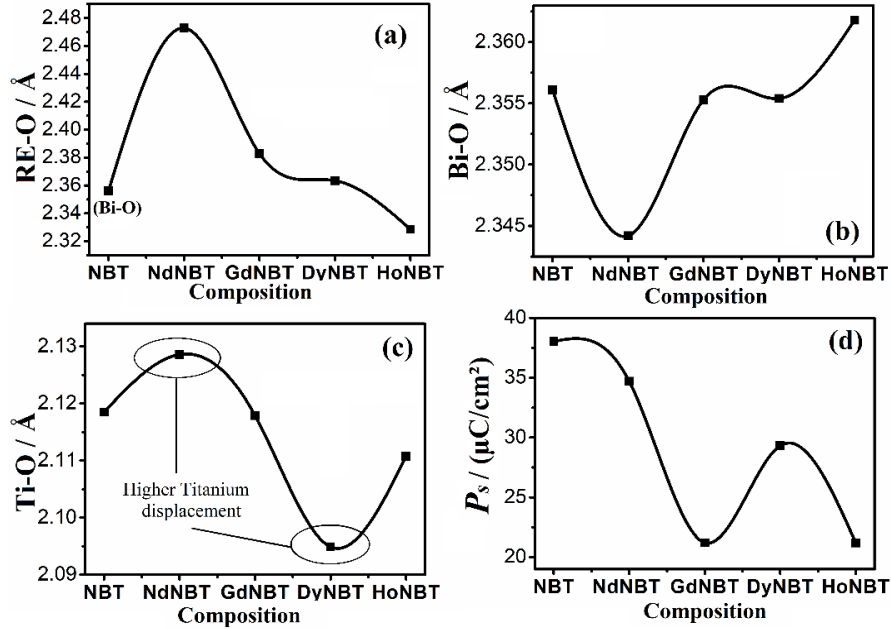

**Fig. 2:** Bond lengths of (a) Rare-earth, (b) bismuth, (c) titanium with oxygen atom for all studied compositions; from high (Nd) to low ionic radii (Ho).

The NdNBT system present the highest polarization value ($P_s$ = 34.7 µC/cm²), which is comparable with some reported experimental values [23,48]. Kandula *et al.* reported a high polarization value in Nd-NBT system which was comparable to the pure NBT material [48]. The observed stronger Bi-O hybridization and higher off-centering of titanium atom, might explain the higher ferroelectric response compared to the other doping. Similarly, a high displacement of titanium atom is observed in the DyNBT compound (Fig. 2(c)), resulting in a relatively high polarization value of $P_s$ = 29.32 µC/cm² and is in good agreement with experimental data [13]. Regarding the GdNBT and HoNBT systems, they present a lower polarization value of ~21.20 µC/cm². Owing to the similar Ti-O and Bi-O hybridization values of NBT, GdNBT, and HoNBT systems, the main parameter affecting the polarization of these last doped compounds would be the evident decrease of the amount of bismuth lone pairs while doping. Note that some papers also reported on the effect of doping NBT-based systems with $Gd^{3+}$ and $Ho^{3+}$ elements [33,49]. For instance, Deng *et al.* showed that a lower piezoelectric coefficient was achieved for $Gd^{3+}$ and $Ho^{3+}$ doped NBT-based system compared to other RE doping [33].

### 3.3. XANES spectra

To further investigate the effect of titanium displacement on the ferroelectric properties of the NBT system, we computed the titanium K-edge x-ray absorption near-edge spectra (XANES) for all studied compositions (Fig. 3(a)). A pre-edge feature is observed on the computed plots (0 – 8 eV), followed by a rising edge (9 – 18 eV). Vedrinskii et al. investigated the nature of the Ti K-edge pre-edge peaks in lead-based ferroelectric PZT system and evidenced that the lower pre-edge peak (Fig. 3(b)) of lower energy (3 eV for NBT) was mainly attributed to the quadrupole transitions of titanium 1s electrons to the unoccupied state of 3d ($t_{2g}$) orbital. In addition, the peak of higher energy (6 eV for NBT) was told to originate from the transition from Ti 1s electron to the absorbing TiO$_6$ polyhedron 3d unoccupied p–d hybrid orbitals of e$_g$ symmetry [50].

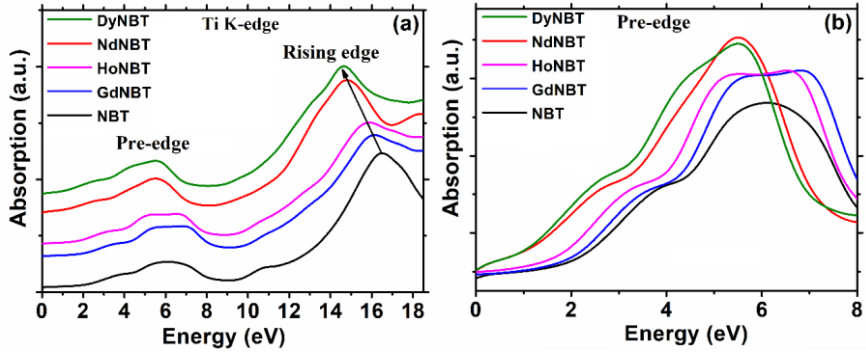

**Fig. 3:** (a) Ti K-edge XANES spectra of RENBT systems, (b) pre-edge feature of Ti K-edge spectra.

Generally, the behavior of the pre-edge feature depends primarily on the structure of the studied compound. For instance, for a centrosymmetric structure, the intensity of the pre-edge can be negligible. Whereas for a non-centrosymmetric structure, it can be significative and is originated from the transition of the metallic 1s electron to the unfilled d orbital [51]. Ravel *et al.* showed the importance of the investigation of the pre-edge feature in ferroelectric perovskites by studying different centrosymmetric (e.g., EuTiO$_3$) and non-centrosymmetric (e.g., BaTiO$_3$) perovskites [51]. Indeed, it can be an efficient tool to probe the magnitude of the octahedral distortions away from centrosymmetry. As observed in Fig. 3(b), the intensity of the absorption spectra is seen to increase first for GdNBT and HoNBT materials, thence further increases for NdNBT and DyNBT systems. Cao *et al.* reported that higher intensity, as well as a higher area under the pre-edge feature of the Ti K-edge, stems for a greater off-centering of Ti atoms from the surrounding oxygen polyhedra [52]. Therefore, in the present study, we believe that a higher off-centering of Ti atom is found in the NdNBT and DyNBT systems, corresponding to the highest absorption intensity. The former observation confirms well the results obtained above. Additionally, a significant shift of the rising edge feature to lower energies for the RENBT compositions is observed (see the arrow in Fig. 3(a)). This last shift is indicative of the decreased charge transfer from the titanium atom in the compounds that are more octahedrally distorted than the NBT system.

### 3.4. Experimental polarization

To confirm the effect of rare-earth doping on the ferroelectric properties of NBT system, we investigated experimentally the polarization of the studied compositions. To do so, ferroelectric hysteresis loops were collected on RENBT ceramics which exhibit pure perovskite phases as shown in Fig. 4(a). The hysteresis loops of the RENBT ceramics illustrated in Fig. 4(b), present typical ferroelectric loops for all compounds. For comparison, the experimental spontaneous polarization is plotted together with theoretical polarization extracted from the BerryPI method in Fig. 4(c). As observed, the trend and the values of the polarization extracted from the BerryPI approach matches well with the one extracted from the hysteresis loops. Notice that the pure NBT system is a non-ergodic relaxor material, i.e., the electric field-induced ferroelectric state is trapped even after the removal of the electric field. The former feature gives rise to a single current (I$_1$) peak for the NBT system (Fig. 4(d)). The introduction of a rare-earth element was found to disturb the ferroelectric domains within the system and instead favors the short-range polar nanoregions (PNRs).

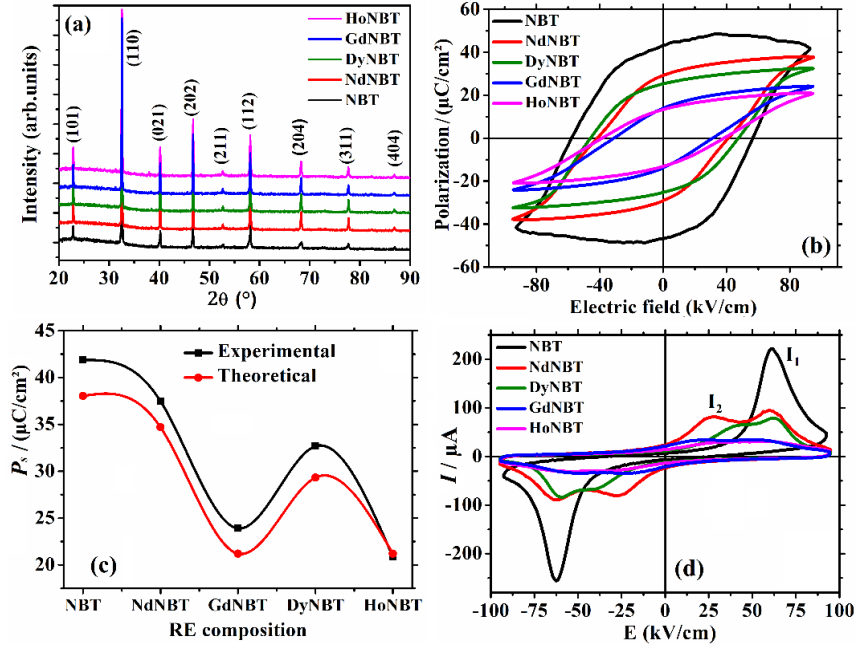

**Fig. 4** (a) X-ray diffraction patterns of the RENBT ceramics. (b) Hysteresis loops of RENBT systems. (c) The maximum polarization extracted from the hysteresis loops, and theoretical polarization extracted from the BerryPI method. (d) The I(E) curves of RENBT systems.

Interestingly, all studied rare-earth-doped NBT systems were observed to present two current peaks, namely, $I_1$ and $I_2$. The later peaks are a signature of the coexistence of both ergodic and non-ergodic relaxor states within the doped systems. The current peak $I_2$ is assigned to the dynamics of the PNRs. Since for an ergodic relaxor material, the electric field-induced long-range FE domains will return to their initial short-range PNRs directly after the removal of the electric field, giving rise to a second current peak, $I_2$. It should be noted that the induced ergodicity is also responsible for the decrease of the remanent polarization, as well as the coercive field values. This effect was observed to be stronger for GdNBT and HoNBT systems showed by their lower $P_r$ and $E_c$ values compared to NdNBT and DyNBT systems. The induced ergodicity within the system is assigned to the increased disorder through the introduction of RE doping element into NBT matrix. To deepen the investigation of the effect of rare-earth on the relaxor properties of the NBT system, we will study in the next section the A-site ordering and relative stability of the pure compound compared to that of a rare-earth-doped NBT system.

### 3.5. A-site Ordering and Relative Stability

Usually ordering or disordering of the AA′ cations of the AA′BO$_3$ perovskite depends on the ionic radii sizes, bonding preferences, and oxidation states of cations [37,47]. Cationic ordering is generally driven by electrostatic considerations due to different ionic radii and oxidation states of cations, whereas cationic disordering is driven by configurational entropy when cations have similar ionic radii and oxidations states. Also, the charge difference ($\Delta q$) value is essential for determining either order ($\Delta q > 2$) or disorder ($\Delta q < 2$) in cations is favored. In the case of $\Delta q = 2$, the arrangement is not resolved and can be fully ordered/disordered or partially ordered [53].

In NBT system, Na$^+$ and Bi$^{3+}$ have a charge difference equal exactly to two ($\Delta q = 2$), in addition to similar ionic radii ($r_{Na+}$ = 1.18 Å, $r_{Bi3+}$ = 1.17 Å). Therefore, full order/disorder or partial order may be favored in the A-site of NBT.

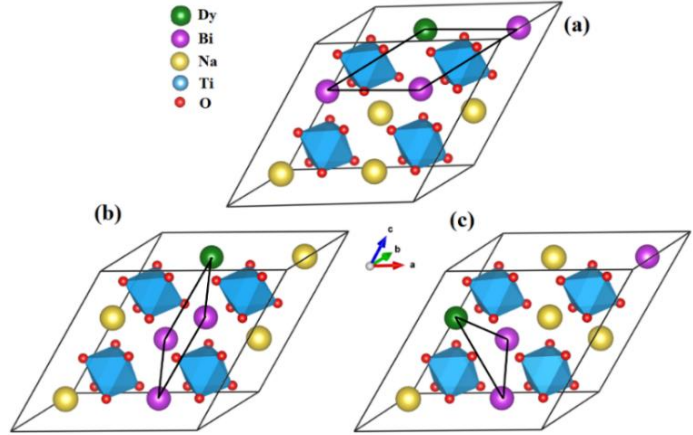

**Fig. 5:** Supercells (2×1×2) with (a) 001, (b) 110, and (c) 111 A-site occupations of DyNBT system.

Hence, we will investigate in this part the relative stability of the different A-site arrangements in two systems, NBT and dysprosium doped NBT material to simulate the effect of rare-earth doping on the chemical ordering. In this aim, we considered three A-site ordering, known as the layered (001), columnar (110), and rock-salt (111) configurations presented in Fig. 5(a-c). We will first investigate the relative stability of the A-site ordering in the pure NBT system, thereafter we will perform an analysis on the case of dysprosium doped NBT system to evaluate the effect of rare-earth introduction on the A-site relative stability.

Notice that, the different A-site arrangements differ in the way the AO$_3$ or A′O$_3$ octahedra connect in the AA′BO$_3$ perovskite (A = Na, A′ = Bi/Dy, B = Ti). Therefore, the layered arrangement allows for the connectivity between the A′O$_3$ octahedra in two dimensions (2D), whereas the columnar arrangement allows only one dimension of connectivity between the A′O$_3$ octahedra (1D). In the case of the rock-salt configuration, there is no connectivity between the A′O$_3$ octahedra in the structure (0D). Generally, B-site ordering is more common than the A-site ordering, which is due to the difference in the oxidation states ($\Delta q$) [54]. In the case of B/B′-site $\Delta q$ may reach seven, whereas in the A/A′-site $\Delta q$ is limited to two or less. Besides, the anion environment differs depending on the A-site arrangement. In the layered arrangement, there are three different anion environments. One where the anion is surrounded by four Na, a second where the anion is surrounded by four Bi, and a third where the anion is surrounded by two Na and two Bi in a cis configuration [55]. The last anion environment is the majoritarian one for the layered arrangement. On the contrary, the rock-salt configuration has only one anion environment, in which the anion is coordinated by two Na and two Bi cations in a trans configuration. Regarding the columnar arrangement, there are three anion environments: two-third having two Bi and two Na in a cis configuration and one third having two Bi and two Na in a trans configuration [40].

**Table 2:** Relative energy ($\Delta E_{relative}$) per formula unit of the different A-site arrangements in the NBT system.

| Compound – A site configuration | NBT – 001 | NBT – 110 | NBT – 111 |
|---|---|---|---|
| $\Delta E_{relative}$ (meV/formula unit) | 0.00 | 440 | 330 |

To investigate the relative stability of the different A-site configurations in the pure NBT system, we computed the energy of each configuration (001, 110, and 111). Therefore, we considered relative energy ($\Delta E_{relative}$), where the 001 direction is the reference ($\Delta E_{relative}$ is obtained by subtracting the energy of the 001-configuration from all compositions) as shown in Table 2. Therefore, by comparing the relative obtained energies, we can conclude that the favored structure i.e., having the lowest relative energy, is the layered arrangement, followed by the rock-salt and finally by columnar A-site order. Generally, the layered order is favored by the A-site, whereas the rock-salt order is favored by the B-site since this ordering maximizes the separation between the highly charged B′ cations allowing the stabilization of the structure [54]. The stabilization of the layered arrangement may be caused by a single environment for all anions. In this last, anions sit in a site with inversion symmetry, and thus the Na-O and Bi-O bonds length remain similar. We agree that the energy difference between each configuration of at least 330 meV/f.u may express a 001-chemical ordering in the NBT matrix. However, the energy difference remains relatively low (meV). This last observation may speak for an A-site ordering (001) that may be favored only in domains, which are trapped in disordered NBT matrix. The former assertion may explain the relaxor properties observed in the NBT system.

To apprehend the effect of rare-earth doping on the chemical ordering of the NBT matrix, we performed a similar investigation in the case of dysprosium doped NBT structure. We computed here too the relative energies per 001 – DyNBT formula unit presented in Table 3. Notice that the lowest energy is given for DyNBT in the layered arrangement with a maximum energy difference of 33 meV/f.u between configurations. It can be observed that the energy difference between the configurations in the case of the doped compound is ten times less than that of the pure matrix. The lower ionic radii of dysprosium compared to that of bismuth result in an increase of the size difference between A and A′ cations in the AA′BO$_3$ perovskite in certain regions favoring an A-site ordering while in other regions the similar ionic radii of Na and Bi will induce a quenched disorder.

**Table 3:** Relative energy ($\Delta E_{relative}$) per formula unit of the different A-site arrangements in DyNBT system.

| Compound – A site configuration | DyNBT – 001 | DyNBT – 110 | DyNBT – 111 |
|---|---|---|---|
| $\Delta E_{relative}$ (meV/formula unit) | 0.00 | 33 | 22 |

Therefore, the weak computed relative energy difference expresses an increased disorder in the system due to the introduction of the rare-earth element. The quenched disorder induced in the doped compound may engender nanoscale structural heterogeneities within the structure. The formation of these heterogeneities, expresses increased relaxor properties in the doped compound compared to the pure system.

### 3.6. Electronic and magnetic properties

To further investigate the effect of the rare earth doping on the electronic properties of NBT system, we computed the total and partial density of electronic states of the RENBT compositions. Total and partial density of electronic states of NBT compound is shown in Fig. S2(a) (see supplementary informations). The Valence Band (VB) is essentially constituted from oxygen (O) 2*p* orbitals, titanate (Ti) 3*d,* and bismuth (Bi) 6*p* orbitals contribution. A strong hybridization occurs between Bi 6*p*, Ti 3*d* orbitals, and O 2*p* orbitals in the VB. Regarding the low energy values of the CB (Conduction Band), the Ti 3*d*, together with Bi 6*p,* are the dominating states. The band structure presented in Fig. S2(b) confirms the electronic bandgap value of the NBT system and permits us to investigate its nature. It is observed that the top of the VB and the bottom of the CB are both situated in the highly symmetric Γ point of the Brillouin zone, which expresses a direct bandgap process.

The effect of rare earth introduction on the electronic properties of the NBT system can be evaluated in Fig. 6(a-d). This last presents the total and partial density of electronic states of the different RENBT systems. The resulted band gap values of all studied compositions extracted from the DOS are gathered in Table 4.

**Table 4**: The electronic band gap, and magnetization *M* ($\mu_B$/f.u) of the different RE doped NBT compositions.

|  | **NBT** | **NdNBT** | **GdNBT** | **DyNBT** | **HoNBT** |
|---|---|---|---|---|---|
| **Band gap (eV)** | 2.56 | 2.55 | 2.57 | 2.57 | 2.54 |
| ***M* ($\mu_B$/f.u)** | 0 | 0.373 | 0.873 | 0.622 | 0.498 |

Remind that, RE elements are characterized by the partially filled 4f orbitals. Each rare earth element has its own electronic configuration, such that, $Nd^{3+}$ ($4f^3$), $Gd^{3+}$ ($4f^7$), $Dy^{3+}$ ($4f^9$), $Ho^{3+}$ ($4f^{10}$). The former 4f orbital is known to split into three-part due to the tetrahedral field: the triply degenerated $t_{1g}$ and $t_{2g}$, in addition to the singly degenerated $a_{2g}$. Fig. 7 presents the degenerated energy levels of the 4f orbital in the case of the different rare earth elements. For instance, in the case of dysprosium element ($4f^9$), the 4f electrons will occupy the majority spin states. Afterward, they will occupy two-thirds of the minority $t_{1g}$ spin states which have the lowest energy. The occupation of the degenerated energy levels of the 4f orbital can be observed in the partial density of the electronic states of the RE 4f orbital given in Fig. 6(a-d). We can observe, that the introduction of the RE ions induced an evident hybridization between the RE 4f and O 2p orbitals. The less filled RE 4f orbital ($4f^3$ of Nd) showed a lower hybridization with oxygen 2p orbitals, in contrary to the other RE elements, and confirms the results obtained in Fig. 2(a). The valence band of the RE 4f orbital present some peaks in the spin up and down channel. For Gd, Dy, and Ho elements, the peaks in the spin up channel are more predominant, owing to their electronic structure (Fig. 7), with 7 up spins filling the three energy levels ($t_{1g}$, $t_{2g}$, and $a_{2g}$), and only few down spins. The conduction band of the RE 4f orbitals can give us informations about the empty energy levels of the 4f orbital (Fig. 7). Neodymium is the only RE element having empty $t_{2g}$, and $a_{2g}$ energy levels which manifest by two peaks in the spin up channel of the CB. Besides, all RE elements present peaks in the spin down channel of the CB, since none of them present completely filled down spins in their energy levels.

The computed total magnetization values of the different RENBT materials in the ground state are gathered in Table 4. As in a nonmagnetic matrix, the total magnetization is principally induced by the 4f orbital of the rare earth elements. It is observed from the table, that the highest magnetization value is obtained in the GdNBT compound ($M = 0.873$ $\mu_B$/f.u). It should be noted that among all considered rare earth elements, the $Gd^{3+}$ ion is the only one possessing seven unpaired spins (Fig. 7), which permits the achievement of a high magnetization in the compound. Regarding the $Dy^{3+}$ and $Ho^{3+}$ doping, all energy levels are filled, however, the $t_{1g}$ energy level is also filled by paired spins which reduces the magnetization in the system to *M*

= 0.622 $\mu_B$/f.u and $M$ = 0.498 $\mu_B$/f.u, respectively. Concerning the $Nd^{3+}$ doping, only the $t_{1g}$ energy level is filled resulting in a lower magnetization value of 0.373 $\mu_B$/f.u.

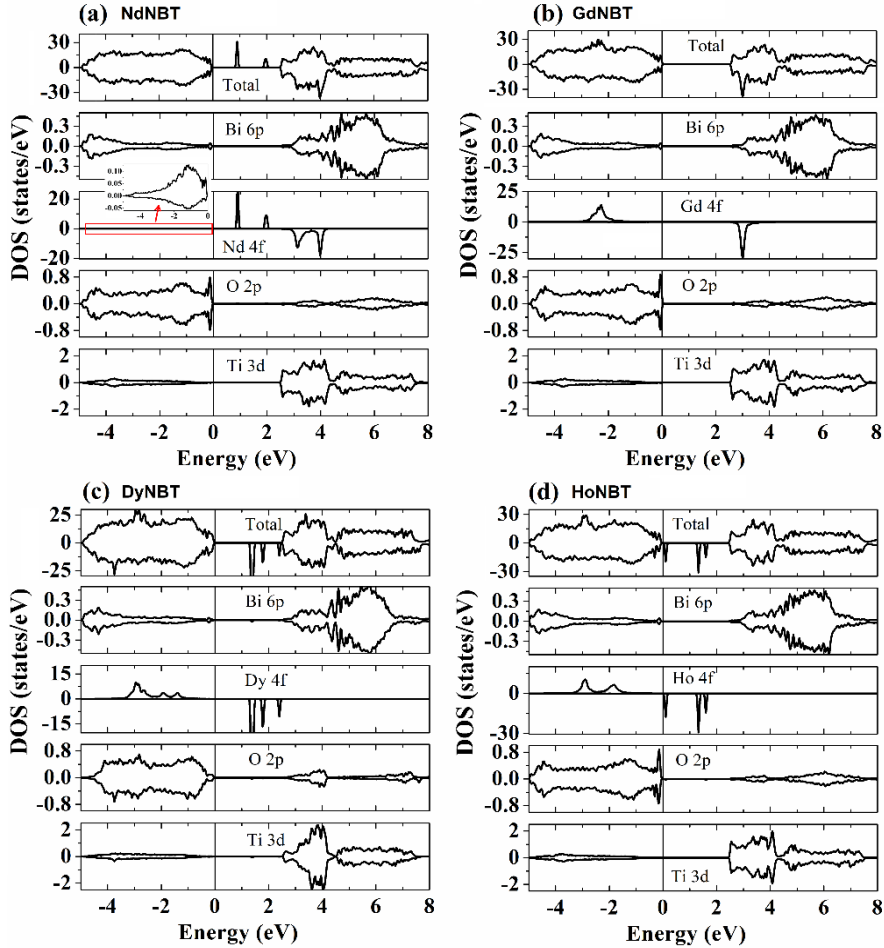

**Fig. 6:** The total and partial density of electronic states of (a) NdNBT, (b) GdNBT, (c) DyNBT, and (d) HoNBT systems.

Note that, experimental investigations on the effect of RE doping on the magnetic properties of NBT system are uncommon. To the best of our knowledge, the only experimental work dealing with this problematic, concerns a paper by Franco Jr. *et al.* [56]. They reported that gadolinium introduction into NBT system increased the room temperature magnetization of the compound, with a transformation from a diamagnetic state to a paramagnetic long-range ordering. However, investigations of the magnetic behavior at high magnetic fields and low temperatures are still missing to understand the magnetic behavior of RE doped NBT systems.

In recent years, interest has been devoted to the introduction of magnetic ions in ferroelectric systems, to create novel multiferroic systems [57]. Thereby, the coexisting magnetic and electric polarization may allow an additional degree of freedom in the design of novel devices (transducers, actuators, storage devices, etc.). In addition to multiple-state memory elements, where data can be stored either by magnetic and electric polarization [58].

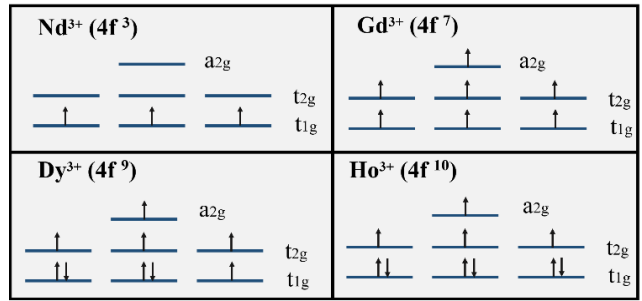

**Fig. 7:** Degenerated energy levels of the 4f orbital of the different rare earth elements.

## 3. Conclusion

In the present work, we performed a comprehensive analysis on the effect of rare-earth (RE = Nd, Gd, Dy, and Ho) doping on the physical properties of the $Na_{0.5}(Bi_{3/4}RE_{1/4})_{0.5}TiO_3$ system using the full potential linearized augmented plane wave based on the spin-polarized density functional theory implemented in the WIEN2k code. The strength of the Bi-O and Ti-O hybridization extracted from the charge density distribution were used to highlight the mechanism responsible for ferroelectricity in the RENBT compounds. The investigation of the Ti K-edge XANES spectra confirmed that the off-centering of titanium from the surrounding oxygen octahedra is principally responsible for the high ferroelectric properties in the NdNBT and DyNBT systems. The trend of the polarization value of the RENBT compounds extracted from the BerryPI method was observed to match very well with the experimental polarization. The electronic investigations showed the emergence of a magnetization owing to the 4f orbital of the RE elements. Likewise, different A-site ordering were investigated and a short-range chemical ordering was highlighted for NBT, and increased A-site disorder was found for DyNBT systems.

# Rare-earth effect on the physical properties of $Na_{0.5}Bi_{0.5}TiO_3$ system: A Density Functional Theory investigation

Manal Benyoussef[1], Halima Zaari[2], Jamal Belhadi[1], Youssef El Amraoui[2,3], Hamid Ez-Zahraouy[2], Abdelilah Lahmar[1], Mimoun El Marssi[1]

[1] Laboratory of Physics of Condensed Matter (LPMC), University of Picardie Jules Verne, Scientific Pole, 80039 Amiens Cedex 1, France
[2] Laboratory of Condensed Matter and Indisciplinary Science (LaMCScI), Faculty of Science, Mohammed V University, BP 1014, Rabat, Morocco
[3] National School of Arts and Crafts, Moulay Ismail University, 50500, Meknes, Morocco

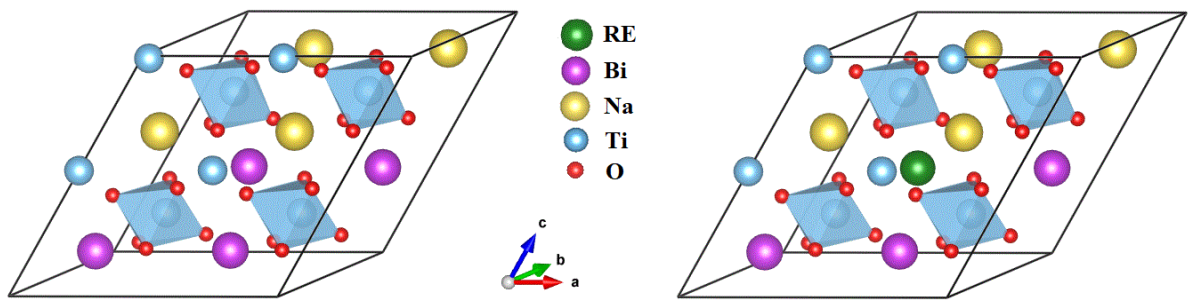

**Fig. S1:** Supercell structure of (a) $Na_{0.5}Bi_{0.5}TiO_3$ system, and (b) $Na_{0.5}(Bi_{3/4}RE_{1/4})_{0.5}TiO_3$

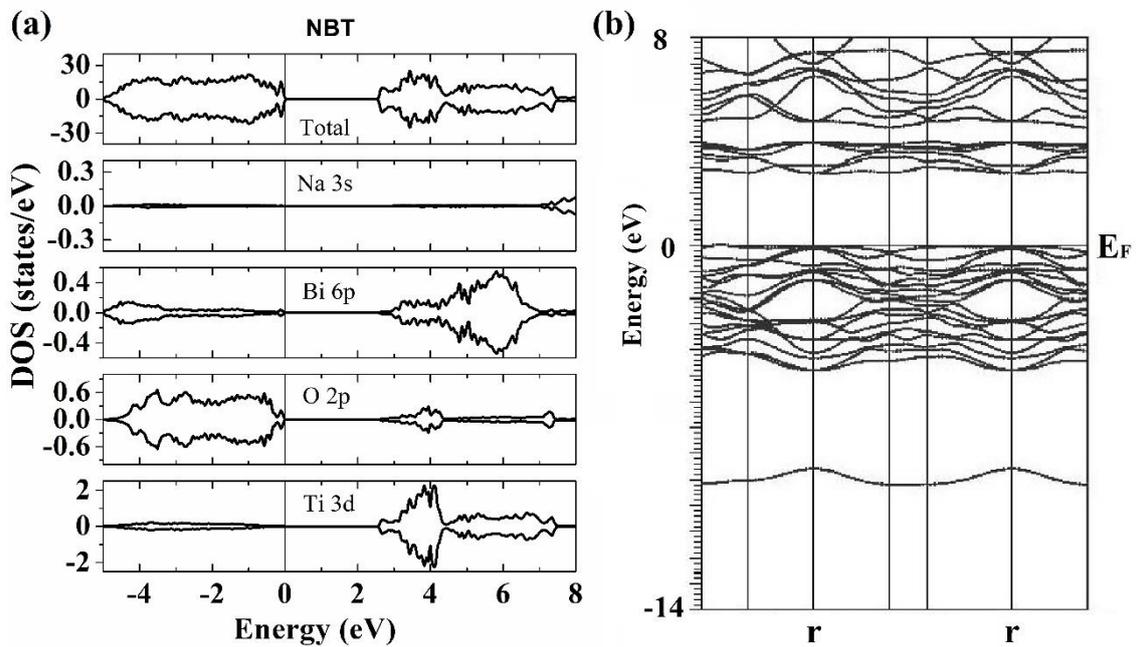

**Fig. S2:** (a) The total and partial density of electronic states, and (b) the Band structure of NBT system.